\documentclass[physrev,reprint,nofootinbib,superscriptaddress]{revtex4-1}

\usepackage{graphicx}
\usepackage{dcolumn}
\usepackage{bm}
\usepackage{amsmath}
\usepackage{amssymb}
\usepackage{booktabs}
\usepackage{multirow}
\usepackage{physics}
\usepackage{xcolor}
\usepackage{pdfpages}

\makeatletter
\AtBeginDocument{\let\LS@rot\@undefined}
\makeatother

\begin{document}

\title{Autonomous Quantum Error Correction in a Four-Photon Kerr Parametric Oscillator}

\author{Sangil Kwon}
\email{kwon2866@gmail.com}
\affiliation{Department of Physics, Tokyo University of Science, Shinjuku, Tokyo 162-0825, Japan}

\author{Shohei Watabe}
\affiliation{Department of Physics, Tokyo University of Science, Shinjuku, Tokyo 162-0825, Japan}

\author{Jaw-Shen Tsai}
\email{tsai@riken.jp}
\affiliation{Department of Physics, Tokyo University of Science, Shinjuku, Tokyo 162-0825, Japan}
\affiliation{RIKEN Center for Quantum Computing (RQC), Wako, Saitama 351-0198, Japan}

\date{\today}

\begin{abstract}
Autonomous quantum error correction has gained considerable attention to avoid complicated measurements and feedback.
Despite its simplicity compared with the conventional measurement-based quantum error correction, it is still a far from practical technique because of significant hardware overhead.
We propose an autonomous quantum error correction scheme for a rotational symmetric bosonic code in a four-photon Kerr parametric oscillator.
Our scheme is the simplest possible error correction scheme that can surpass the break-even point---it requires only a single continuous microwave tone.
We also introduce an unconditional reset scheme that requires one more continuous microwave tone in addition to that for the error correction.
The key properties underlying this simplicity are protected quasienergy states of a four-photon Kerr parametric oscillator and the degeneracy in its quasienergy level structure.
These properties eliminate the need for state-by-state correction in the Fock basis.
Our schemes greatly reduce the complexity of autonomous quantum error correction and thus may accelerate the use of the bosonic code for practical quantum computation.
\end{abstract}

\maketitle

%\tableofcontents

\section{Introduction}

The most serious obstacle towards fault-tolerant quantum computation is probably quantum error correction.
The reason is that quantum error correction requires a large Hilbert space as well as high-fidelity measurement and control.
The use of a harmonic oscillator, i.e., a bosonic system, is one strategy to obtain a large Hilbert space without too much hardware overhead \cite{terhal2020, cai2021, joshi2021, ma2021}.
In this system, information can be encoded as a symmetric pattern in phase space.
Such a symmetry can be either translational (Gottesman--Kitaev--Preskill code) \cite{grimsmo2021, gkp} or rotational (cat or binomial code) \cite{grimsmo2020}.
In a superconducting circuit \cite{cQED, kwon, mit, gu, tsai}, which is our working system, another advantage of using a harmonic oscillator is that the major loss mechanism is single-photon loss;
thus, quantum error correction in this system can be achieved by detecting the number parity of the photon state \cite{sun2014, ofek2016, hu2019}.

Recently, autonomous quantum error correction (AQEC) schemes have gained considerable attention to avoid complicated measurements and feedback \cite{sarovar2005, leghtas2013, mirrahimi2014, cohen2014, kapit2016, lihm2018, albert2019, ma2020, gertler2021, wang2021}.
Although AQEC is considered to be much easier to implement than the conventional measurement-based quantum error correction, there is still a serious hardware overhead---the need for many microwave tones.
The origin of this problem is that the logical qubit states are composed of multiple Fock states, and errors are corrected by selective transitions induced by continuous microwave tones.
Since each transition requires a separate continuous microwave tone, many microwave tones are required to handle all possible transitions.
For example, eight microwave tones are used in Ref.~\cite{gertler2021} although the logical qubit states in this reference are composed of only four Fock states.
Moreover, the amplitude of each tone must be tuned independently to ensure identical transition rates that prevent leakage of which-path information.
Thus, any scheme based on state-by-state correction in the Fock basis is difficult to scale up.

In this study, we propose an AQEC scheme that requires only a single continuous microwave tone---the simplest possible error correction scheme that can surpass the break-even point.
This substantial reduction of hardware overhead is due to the protection of the Hilbert space for encoding and error correction such that the system remains in this protected Hilbert space under single-photon loss/drive.
Such a protection is provided by a four-photon pump applied to a Kerr nonlinear oscillator---a system with a small anharmonicity of less than 1\% of its resonance frequency \cite{wustmann2019}.
Since this four-photon pump cannot be achieved by simple linear driving and must be achieved by parametric modulation of the Josephson junction energy \cite{svensson2017, svensson2018} (see Supplementary Note 1), we term this system a four-photon Kerr parametric oscillator (KPO).
Although a KPO has received much attention very recently because of its use for the generation and stabilization of the cat states \cite{guo2020, goto2019b, guo2013, goto2016a, minganti2016, puri2017a, zhang2017, masuda2021, wang2019, grimm2020},
gate-based quantum computation \cite{goto2019b, puri2017a, grimm2020, goto2016b, puri2020, kanao2021b, xu2021}, 
measurement-based error correction \cite{puri2019, darmawan2021},
quantum annealing \cite{nigg2017, puri2017b, zhao2018, goto2019c, onodera2020, goto2020, kanao2021a}, and
other physically interesting topics \cite{goto2019a, strandberg2021, goto2018, mamaev2018, kewming2020, savona2017, rota2019, goto2021},
little attention has been paid to its applicability to AQEC.
Our study reveals that a KPO can be a suitable system for AQEC.

\section{Results}

\subsection{System and encoding}

%%%%%%%%%%%%%%%%%%%%%%%%%%%%%%%%
\begin{figure*}
\centering
\includegraphics{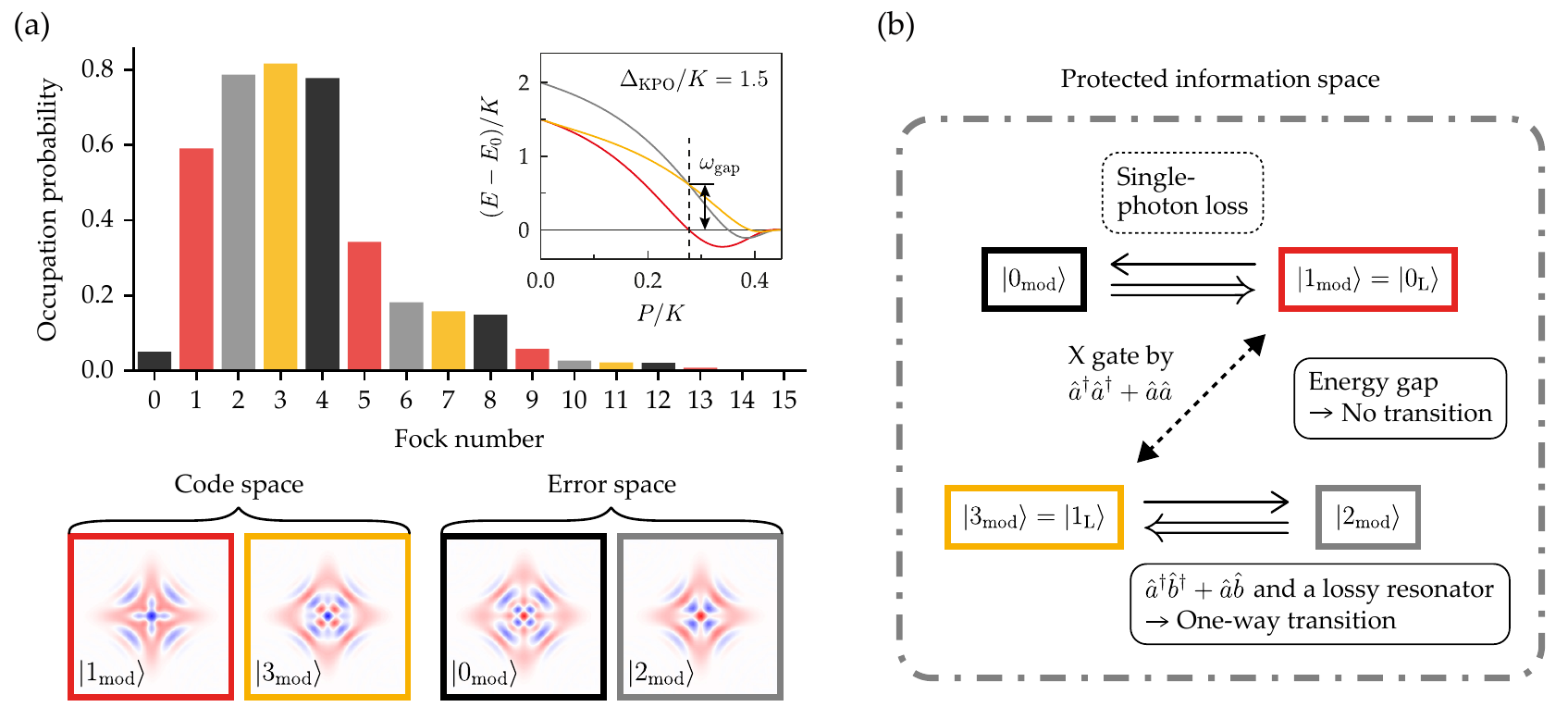}
\caption{Information encoding and AQEC scheme in a four-photon KPO.
(a) Four eigenstates in the information space.
The upper panel shows the occupation probability of these eigenstates in the Fock basis at $\Delta_\textrm{KPO}/K = 1.5$ and $P/K = 0.2764$ (indicated by a vertical dashed line in the inset), where the energy levels of $\ket{0_\textrm{mod}}$ and $\ket{1_\textrm{mod}}$ ($\ket{2_\textrm{mod}}$ and $\ket{3_\textrm{mod}}$) are degenerated.
Here, $E_0$ is the energy level of $\ket{0_\textrm{mod}}$.
The inset shows the quasienergy levels of the four eigenstates as a function of $P$.
The four lower panels show the Wigner distribution of the four eigenstates.
The eigenstates with the odd/even number parity consist of the code/error space.
(b) Requirements for AQEC.
The dash-dot line forming the gray boundary represents protection of the information space provided by the four-photon pump.
Double arrows ($\Rightarrow$) indicate induced transitions, and single arrows ($\rightarrow$) indicate spontaneous transitions caused by single-photon loss.
Colors of bars and frames in this figure indicate the modulus of 4 in the Fock basis.
}
\label{fig:protect}
\end{figure*}
%%%%%%%%%%%%%%%%%%%%%%%%%%%%%%%%

Our system of interest is a KPO driven by a four-photon pump whose frequency is $\omega_\textrm{p}$.
In the rotating frame with the frequency $\omega_\textrm{p}/4$, the Hamiltonian of the KPO is given by (see Sec.~\ref{sec:gate} and Supplementary Note 1 for circuit implementation and derivation of the Hamiltonian)
\begin{equation}\label{eq:KPO}
\begin{split}
\hat{\mathcal{H}}_\textrm{KPO} &=
\hbar\Delta_\textrm{KPO} \hat{a}^\dagger\hat{a}
- \hbar\frac{K}{2} \hat{a}^\dagger\hat{a}^\dagger \hat{a}\hat{a} \\
&\quad
+ \hbar\frac{P}{2} \left( \hat{a}^\dagger\hat{a}^\dagger\hat{a}^\dagger\hat{a}^\dagger + \hat{a}\hat{a}\hat{a}\hat{a} \right).
\end{split}
\end{equation}
Here,
$\hat{a}$ and $\hat{a}^\dagger$ are the ladder operators for the KPO,
$\Delta_\textrm{KPO} (\equiv \omega_\textrm{KPO} - \omega_\textrm{p}/4)$ is the KPO-pump frequency detuning, where $\omega_\textrm{KPO}$ is the transition frequency of the KPO between $\ket{0}$ and $\ket{1}$ states,
$K$ is the Kerr coefficient, and
$P$ is the amplitude of the pump.

The four highest quasienergy states from Eq.~\eqref{eq:KPO} are energetically close and show fourfold rotational symmetry in phase space as shown in Fig.~\ref{fig:protect}(a).
These states are represented by the modulus of 4 in the Fock basis:
\begin{align}
\ket{k_\textrm{mod}} = \sum_{n=0}^\infty C_n^{(k)}\ket{4n+k},
\end{align}
where $\big|C_n^{(k)}\big|^2$ ($k\in \{0,1,2,3\}$) indicates the occupation probability of each Fock basis, which is plotted in the upper panel of Fig.~\ref{fig:protect}(a).

We encode information on the states with the odd number parity \cite{gertler2021}---$\ket{1_\textrm{mod}}$ as $\ket{0_\textrm{L}}$ and $\ket{3_\textrm{mod}}$ as $\ket{1_\textrm{L}}$, where the subscript L denotes the logical qubit states.
These two logical states comprise the code space (warm colors in Fig.~\ref{fig:protect}), whereas the remaining two states with even number parity constitute the error space (achromatic colors).
In this work, we call the code space and error space together the information space [the gray dash-dot boundary in Fig.~\ref{fig:protect}(b)].

Note that there is an energy gap, which we call the protection energy gap, the size of which is about $3K$ at $\Delta_\textrm{KPO}$ and $P$ values shown in the caption of Fig.~\ref{fig:protect}(a).
This energy gap protects the information space by suppressing the population leakage to states outside of the information space, which we name `higher excitation levels (HEL)' although their quasienergies are actually lower because of the minus sign in front of $K$ [Eq.~\eqref{eq:KPO}].
(The quasienergy level diagram showing the protection energy gap and the HEL space are shown in Supplementary Figure 4.)

We can access $\ket{1_\textrm{mod}}$ and $\ket{3_\textrm{mod}}$ by increasing $P$ in Eq.~\eqref{eq:KPO} adiabatically from $\ket{1}$ and $\ket{3}$, respectively \cite{goto2016a, puri2017a}.
In Sec.~\ref{sec:reset}, we discuss another convenient way to reset the state of the system to the logical qubit states unconditionally by applying two continuous microwave tones.

\subsection{Error correction scheme}

Our autonomous error correction scheme is shown in Fig.~\ref{fig:protect}(b).
Our scheme corrects errors caused by single-photon loss and thus relies on change in the number parity of the KPO state \cite{mirrahimi2014}.
The crucial observation is that if $\ket{1_\textrm{mod}}$ ($\ket{3_\textrm{mod}}$) loses one photon, the final state is likely $\ket{0_\textrm{mod}}$ ($\ket{2_\textrm{mod}}$) because of the protection by the four-photon pump.
This means that we can recover $\ket{1_\textrm{mod}}$ from $\ket{0_\textrm{mod}}$ ($\ket{3_\textrm{mod}}$ from $\ket{2_\textrm{mod}}$) by applying a single-photon drive.
Note that we cannot ask which Fock state loses or gains the photon as all Fock states comprising the logical qubit state change simultaneously.
This eliminates the need to control the Fock states one by one, thus greatly reducing hardware overhead.

The second essential requirement, other than the protection of the information space, is the energy degeneracy between $\ket{0_\textrm{mod}}$ and $\ket{1_\textrm{mod}}$ as well as between $\ket{2_\textrm{mod}}$ and $\ket{3_\textrm{mod}}$.
Remarkably, this can be achieved simply by tuning $P$ in Eq.~\eqref{eq:KPO} to the value indicated by the vertical dashed line in the inset of Fig.~\ref{fig:protect}(a).
The energy degeneracy allows us to induce transitions between $\ket{0_\textrm{mod}}$ and $\ket{1_\textrm{mod}}$ as well as between $\ket{2_\textrm{mod}}$ and $\ket{3_\textrm{mod}}$ using a single microwave tone with the frequency $\omega_\textrm{p}/4$.

Other requirements for our AQEC scheme are
(i) one-way transition: $\ket{0_\textrm{mod}} \rightarrow \ket{1_\textrm{mod}}$ and $\ket{2_\textrm{mod}} \rightarrow \ket{3_\textrm{mod}}$, and
(ii) no transition: $\ket{1_\textrm{mod}} \nleftrightarrow \ket{2_\textrm{mod}}$ and $\ket{3_\textrm{mod}} \nleftrightarrow \ket{0_\textrm{mod}}$.
The one-way transition can be realized by introducing an ancilla resonator  whose ladder operators are $\hat{b}^\dagger$ and $\hat{b}$, and applying $\hat{a}^\dagger\hat{b}^\dagger + \hat{a}\hat{b}$ instead of $\hat{a}^\dagger+\hat{a}$.
(Hereafter, we refer to the microwave tone for $\hat{a}^\dagger\hat{b}^\dagger + \hat{a}\hat{b}$ as the correction tone.)
This ancilla resonator must be very lossy compared with the KPO to suppress transitions from the code space to the error space.
The resulting process is as follows.
For example, $\ket{0_\textrm{mod}}$ can be corrected with the ancilla, whose state is written with the subscript `an', as
\begin{equation}
\ket{0_\textrm{mod}, 0_\textrm{an}} \Rightarrow \ket{1_\textrm{mod}, 1_\textrm{an}} \rightarrow \ket{1_\textrm{mod}, 0_\textrm{an}},
\end{equation}
where $\Rightarrow$ indicates a transition induced by the correction tone, whereas $\rightarrow$ indicates a spontaneous transition in the lossy ancilla.
Transitions such as $\ket{1_\textrm{mod}} \leftrightarrow \ket{2_\textrm{mod}}$ and $\ket{3_\textrm{mod}} \leftrightarrow \ket{0_\textrm{mod}}$ can be suppressed by an energy gap $\omega_\textrm{gap}$ (different from the protection energy gap).
Note that we have this energy gap already---see the inset of Fig.~\ref{fig:protect}(a).

\subsection{Numerical simulation}
\label{sec:sim}

%%%%%%%%%%%%%%%%%%%%%%%%%%%%%%%%
\begin{figure*}
\centering
\includegraphics{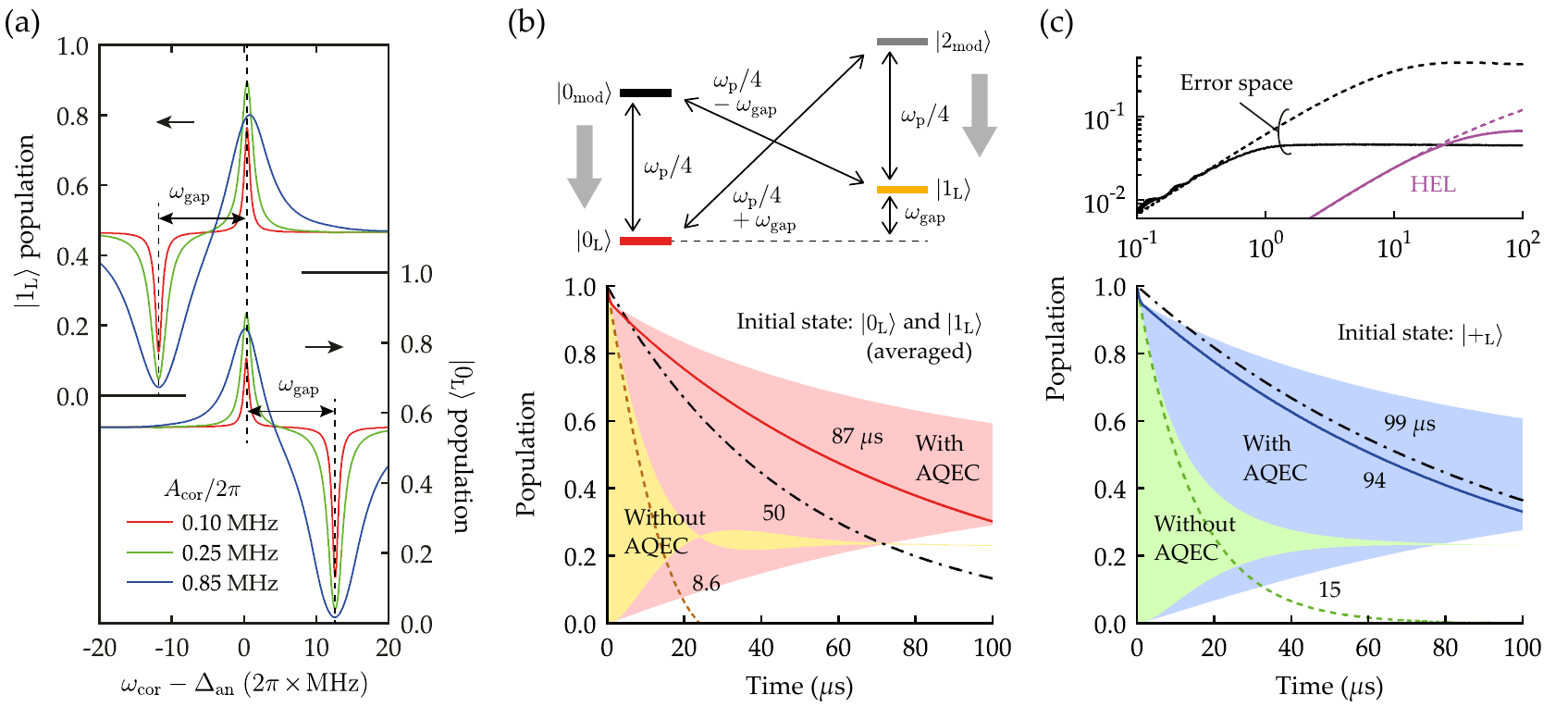}
\caption{Autonomous error correction.
(a) Population of the logical qubit states as a function of $\omega_\textrm{cor}$ after 10 $\mu$s of evolution with various $A_\textrm{cor}$.
The distance between the peak and the dip is $\omega_\textrm{gap}$, which is found to be 12.2 MHz.
%%%
(b) Lower panel: Bit-flip process of the logical qubit states with and without AQEC.
The upper (lower) envelop of the shaded area is the population of the initial logical state (the orthogonal logical state).
The red solid and brown dashed lines indicate the population difference between the upper and lower envelops with and without AQEC, respectively.
The number near each curve indicates the corresponding bit-flip time extracted from exponential fitting.
The black dash-dot line indicates the longitudinal relaxation in the $\ket{0}$ and $\ket{1}$ encoding.
Upper panel: Frequencies of the microwave tone (not the energy levels) at which transitions are induced among the four eigenstates.
The thick gray arrows constitute a graphical summary of our AQEC scheme.
%%%
(c) Lower panel: Phase-flip process of the logical qubit states with and without AQEC.
The number near each curve indicates the corresponding phase-flip time.
The black dash-dot line indicates the transverse relaxation in the $\ket{0}$ and $\ket{1}$ encoding.
Other initial states, $\ket{-_\textrm{L}}$, $\ket{\textrm{i}+_\textrm{L}}$, and $\ket{\textrm{i}-_\textrm{L}}$, where $\ket{\pm_\textrm{L}} \equiv (\ket{0_\textrm{L}}\pm\ket{1_\textrm{L}})/\sqrt{2}$ and $\ket{\textrm{i}\pm_\textrm{L}} \equiv (\ket{0_\textrm{L}}\pm\textrm{i}\ket{1_\textrm{L}})/\sqrt{2}$, present identical results.
Upper panel: Population leakage to the error and HEL spaces during the evolution from $\ket{+_\textrm{L}}$.
The solid and dashed lines represent the populations with and without AQEC, respectively.
Other initial states, $\ket{0_\textrm{L}}$ and $\ket{1_\textrm{L}}$, yield similar results.
%%%
The parameters are as follows.
KPO:
$\omega_\textrm{KPO}/2\pi = 2.98$ GHz,
$K/2\pi = 20$ MHz,
$1/\gamma_\textrm{KPO} = 50$ $\mu$s \cite{osman2021}.
Pump:
$\Delta_\textrm{KPO}/2\pi = 30$ MHz,
$P/2\pi = 5.5405$ MHz.
Ancilla resonator: 
$\omega_\textrm{an}/2\pi = 4$ GHz
$\gamma_\textrm{an}/2\pi = 0.557$ MHz [found in Fig.~\ref{fig:opt}(c)],
$g/2\pi = 7$ MHz.
Correction tone:
$\omega_\textrm{cor}/2\pi = \Delta_\textrm{an}/2\pi + 0.36$ MHz [found in Fig.~\ref{fig:opt}(a)], and
$A_\textrm{cor}/2\pi = 0.25$ MHz [found in Fig.~\ref{fig:opt}(b)].
}
\label{fig:AQEC}
\end{figure*}
%%%%%%%%%%%%%%%%%%%%%%%%%%%%%%%%

We simulate our AQEC scheme by solving the following master equation \cite{puri2019} with QuTiP \cite{qutip1, qutip2}.
\begin{equation}\label{eq:lindblad}
\begin{split}
\frac{\partial \rho(t)}{\partial t} &= 
-\frac{\textrm{i}}{\hbar} [\hat{\mathcal{H}}_\textrm{full}(t),\rho(t)] \\
&\quad
+ \big\{
\gamma_\textrm{KPO}(1+n_\textrm{th})\mathcal{D}[\hat{a}]
+ \gamma_\textrm{KPO} n_\textrm{th} \mathcal{D}[\hat{a}^\dagger] \\
&\quad
+ \gamma_\phi\mathcal{D}[\hat{a}^\dagger\hat{a}]
+ \gamma_\textrm{an}\mathcal{D}[\hat{b}]
\big\}\rho(t),
\end{split}
\end{equation}
where $\mathcal{D}[\hat{O}]\rho = \hat{O}\rho\hat{O}^\dagger - \frac{1}{2}\hat{O}^\dagger\hat{O}\rho - \frac{1}{2}\rho\hat{O}^\dagger\hat{O}$,
$\gamma_\textrm{KPO}$ is the single-photon loss rate of the KPO,
$n_\textrm{th}$ is the number of thermal photons in the KPO,
$\gamma_\phi$ is the dephasing rate of the KPO, and
$\gamma_\textrm{an}$ is the single-photon loss rate of the ancilla resonator.
In this subsection, we consider only single-photon losses in the KPO and the ancilla resonator.
(The effects of $n_\textrm{th}$ and $\gamma_\phi$ will be discussed in Sec.~\ref{sec:relax}.)
The time-dependent Hamiltonian $\hat{\mathcal{H}}_\textrm{full}(t)$ is given by
\begin{equation}\label{eq:full}
\begin{split}
\hat{\mathcal{H}}_\textrm{full}(t) &\approx
\hat{\mathcal{H}}_\textrm{KPO} 
+ \hbar \Delta_\textrm{an} \hat{b}^\dagger\hat{b} 
+ \hbar g(\hat{a}^\dagger\hat{b} + \hat{a}\hat{b}^\dagger) \\
&\quad
+ \hbar A_\textrm{cor}\cos(\omega_\textrm{cor}t)(\hat{a}^\dagger\hat{b}^\dagger+\hat{a}\hat{b}).
\end{split}
\end{equation}
Here, 
$\Delta_\textrm{an} (\equiv \omega_\textrm{an} - \omega_\textrm{p}/4)$ is the ancilla-pump frequency detuning, where $\omega_\textrm{an}$ is the resonance frequency of the ancilla resonator,
$g$ is the coupling constant between the KPO and the ancilla, and
$A_\textrm{cor}$ and $\omega_\textrm{cor}$ are the amplitude and the frequency of the AQEC term, respectively.

The first thing that must be done for AQEC is to find the appropriate $\omega_\textrm{cor}$.
The result of such a frequency sweep is shown in Fig.~\ref{fig:AQEC}(a).
We found a peak in the population of the logical qubit states for $\omega_\textrm{cor}$ near $\Delta_\textrm{an}$---a signature of AQEC.
This result is consistent with the $\hat{a}^\dagger\hat{b}^\dagger+\hat{a}\hat{b}$ term because $\Delta_\textrm{an}$ corresponds to $\omega_\textrm{an}+\omega_\textrm{p}/4$ in the lab frame.
In addition, we found a dip separated from the peak with $\pm \omega_\textrm{gap}$ whose sign depends on the logical qubit state of interest.
This dip activates the transitions suppressed by $\omega_\textrm{gap}$.
Such a peak-dip structure can be understood with the diagram shown in the upper panel of Fig.~\ref{fig:AQEC}(b).
In this diagram, our AQEC scheme can be understood as population transfer along the thick gray arrows.

Our main results, the relaxation times with and without AQEC, are presented in Fig.~\ref{fig:AQEC}(b) and (c).
With optimal $A_\textrm{cor}$ and $\gamma_\textrm{an}$ values (the optimization procedure for these quantities will be discussed in Sec.~\ref{sec:opt}),
the bit-flip time of the logical qubit states is increased by approximately one order of magnitude [the lower panel of Fig.~\ref{fig:AQEC}(b)],
and the phase-flip time is increased by over a factor of 6 [the lower panel of Fig.~\ref{fig:AQEC}(c)].
Note that the phase-flip time is not greater than $T_2$ in $\ket{0}$ and $\ket{1}$ encoding.
This limited performance of AQEC is likely due to different mean photon number between two logical qubit states (see Sec.~\ref{sec:opt} for further discussion).
The resulting relaxation time of the process fidelity \cite{ofek2016} surpasses the break-even point by approximately 20\%.
We believe that surpassing break-even point will not be too difficult in experiments.
The reason is that dephasing due to the low-frequency noise \cite{kwon} was not considered in our simulation---that is $T_2 = 2T_1$ in $\ket{0}$ and $\ket{1}$ encoding, where $T_2$ and $T_1$ are the transverse and longitudinal relaxation times, respectively---whereas all planar superconducting circuits are sensitive to low frequency noise, such that $T_2$ is often significantly less than $2T_1$.
Note that our encoding in a four-photon KPO is insensitive to such a noise.
This is because the collapse operator that models the dephasing process, $\sqrt{\gamma_\phi}\hat{a}^\dagger\hat{a}$, induces population leakage out of the information space [see Sec.~\ref{sec:relax} and Fig.~\ref{fig:relax}(c)] and this process requires an energy greater than the protection energy gap \cite{puri2017a, nigg2017, puri2019, grimm2020}.

Another relaxation process other than bit and phase flips is population leakage out of the code space.
The origin of population leakage to the HEL space is finite transition probability between the code and HEL spaces---for example, 
$\big|\!\mel{0_\textrm{mod}^\textrm{h}}{\hat{a}}{1_\textrm{mod}}\!\big|^2 = 0.073$ and 
$\big|\!\mel{2_\textrm{mod}^\textrm{h}}{\hat{a}}{3_\textrm{mod}}\!\big|^2 = 0.029$---where the lowercase h indicates that the state is a part of the HEL space.
By calculating the population of all states, including the HEL space (see Supplementary Table 1), we find that our AQEC scheme is effective in reducing the population of the error space, which is suppressed by more than one order of magnitude after 10 $\mu$s [the upper panel of Fig.~\ref{fig:AQEC}(c)].
The population of the HEL space is also suppressed by 44\% at 100 $\mu$s.
This population leakage has been called quantum heating---a heating process induced by quantum jumps due to dissipation (in this case, single-photon loss) in quasienergy levels of driven quantum nonlinear systems \cite{goto2018, marthaler2006, dykman2011, ong2013}.
The steady state solution in Supplementary Table 1 is indeed independent of $\gamma_\textrm{KPO}$, which is a signature of quantum heating \cite{goto2018, dykman2011}.
The AQEC process does not generate quantum heat because the correction tone is weak and continuous; thus, transitions over the protection gap do not occur.
Thus, it can be said that AQEC cools quantum heat.

\subsection{Optimization}
\label{sec:opt}

%%%%%%%%%%%%%%%%%%%%%%%%%%%%%%%%
\begin{figure}
\centering
\includegraphics{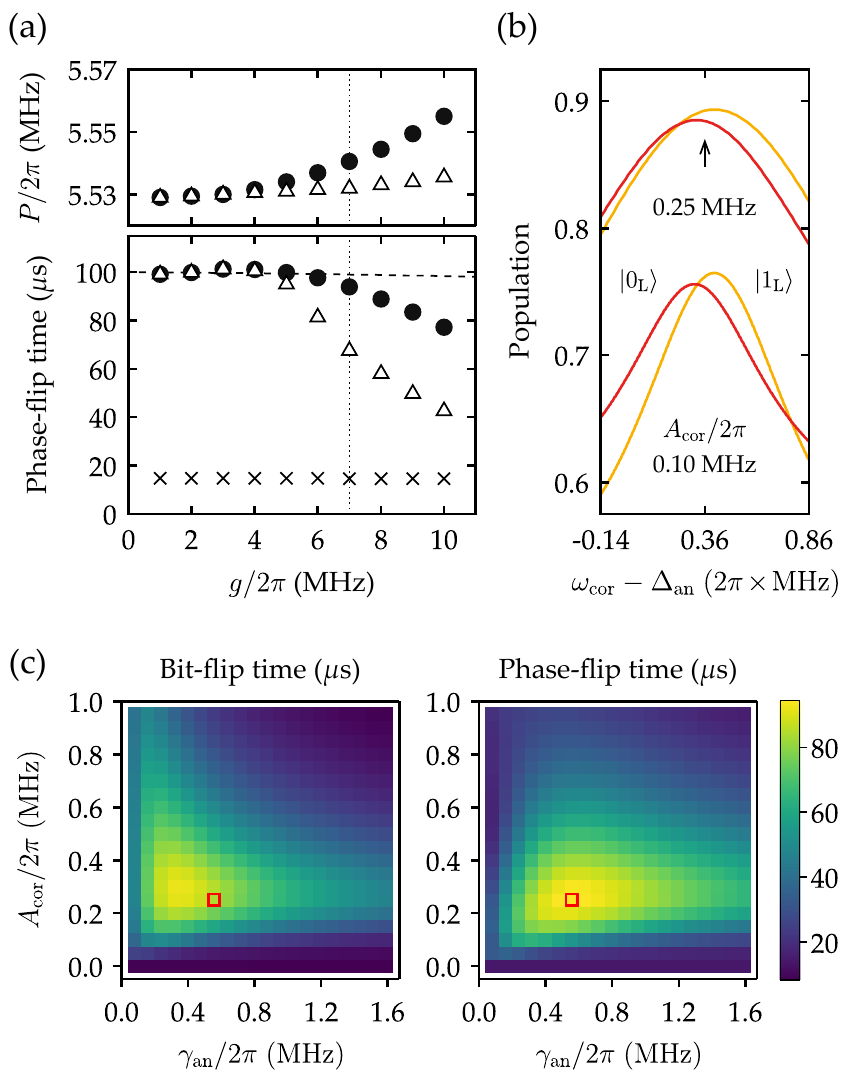}
\caption{Optimization procedure.
(a) Upper panel: Four-photon pump amplitude as a function of the KPO-ancilla coupling constant.
The solid circle indicates the pump amplitude where the phase-flip time is maximized and the empty triangle indicates the pump amplitude where quasienergy levels degenerate.
Lower panel: Phase-flip time with two different four-photon pump amplitudes.
The symbols are the same as those in the upper panel, except the cross symbol indicates the phase-flip time without AQEC.
The dashed horizontal line indicates $T_2$ in the $\ket{0}$ and $\ket{1}$ encoding.
The dotted vertical line indicates the coupling constant that we used in this work, $g/2\pi=7$ MHz.
(b) Populations of the logical qubit states as a function of $\omega_\textrm{cor}$ after 10 $\mu$s of evolution with $A_\textrm{cor}/2\pi = 0.10$ and 0.25 MHz.
The arrow indicates the frequency we used as the optimal $\omega_\textrm{cor}$.
(c) Bit- and phase-flip times as a function of $A_\textrm{cor}$ and $\gamma_\textrm{an}$ at the optimal $\omega_\textrm{cor}$.
The red squares, where the phase-flip time is maximized, indicate the conditions used in Fig.~\ref{fig:AQEC}.
}
\label{fig:opt}
\end{figure}
%%%%%%%%%%%%%%%%%%%%%%%%%%%%%%%%

A potential problem of our encoding is that the mean photon number of $\ket{0_\textrm{L}}$ and $\ket{1_\textrm{L}}$ are 2.9 and 3.8, respectively, thereby suggesting that our logical qubit does not satisfy Knill--Laflamme conditions \cite{NC, BCRS}.
One consequence of this is that the probabilities of single-photon loss events in the two logical qubit states are different, thereby resulting in information leakage directly to the environment or indirectly via the ancilla resonator.
The information leakage path via the ancilla can be minimized by designing the dispersive shift due to the coupling between the KPO and the ancilla being much smaller than the linewidth of the ancilla.
Simultaneously, $g$ must be sufficiently large to generate a reasonably high $A_\textrm{cor}$ from the correction tone because $A_\textrm{cor}$ is determined by $g$, although these two are written as independent parameters in Eq.~\eqref{eq:full}.
We find that $g=7$ MHz used in Fig.~\ref{fig:AQEC} meets these criteria (see the end of Supplementary Note 1).

Remarkably, the phase-flip time increases significantly when the four-photon pump amplitude $P$ is slightly higher than the value for energy degeneracy, when $g > 4$ MHz [Fig.~\ref{fig:opt}(a)].
This slight detuning separates the population peaks of $\ket{0_\textrm{L}}$ and $\ket{1_\textrm{L}}$ as depicted in Fig.~\ref{fig:opt}(b).
We set the frequency at the center of two peaks as the optimal $\omega_\textrm{cor}$.

Other parameters, $A_\textrm{cor}$ and $\gamma_\textrm{an}$, can be optimized to maximize the bit- and phase-flip times by sweeping the parameter space [Fig.~\ref{fig:opt}(c)].
The reason for existence of the optimal $A_\textrm{cor}$ is that if $A_\textrm{cor}$ is too large, the height of the peak decreases because the dip becomes broader and eventually undermines the peak, as presented in Fig.~\ref{fig:AQEC}(a).

\subsection{Photon gain and dephasing}
\label{sec:relax}

%%%%%%%%%%%%%%%%%%%%%%%%%%%%%%%%
\begin{figure}
\centering
\includegraphics{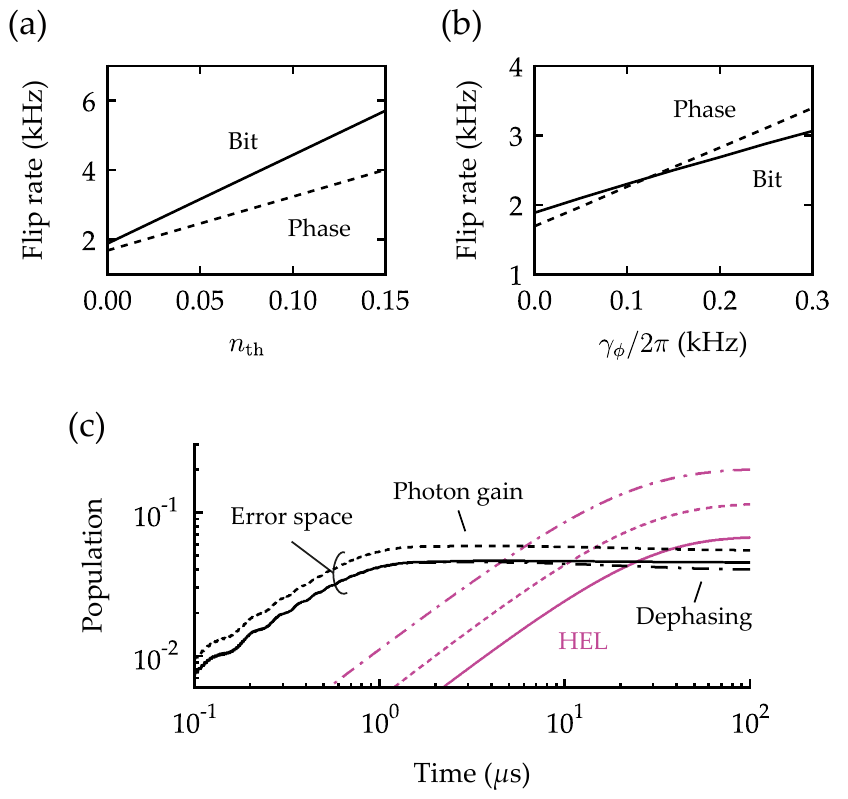}
\caption{Effects of single-photon gain and dephasing.
(a) and (b) Bit- and phase-flip times as a function of single-photon gain ($n_\textrm{th}$) and dephasing ($\gamma_\phi$).
The bit- and phase-flip times are obtained via the same procedure depicted in Fig.~\ref{fig:AQEC}(b) and (c).
(c) Population leakage to the error and HEL spaces with AQEC in the presence of single-photon gain and effective dephasing.
The solid lines indicate the populations when $n_\textrm{th} = 0$ and $\gamma_\phi/2\pi = 0$ kHz;
the dashed lines, $n_\textrm{th} = 0.15$ and $\gamma_\phi/2\pi = 0$ kHz;
the dash-dot lines, $n_\textrm{th} = 0$ and $\gamma_\phi/2\pi = 0.3$ kHz.
The result in (c) is the average of two evolutions from $\ket{0_\textrm{L}}$ and $\ket{1_\textrm{L}}$;
the evolution from $\ket{+_\textrm{L}}$ yields similar results.
}
\label{fig:relax}
\end{figure}
%%%%%%%%%%%%%%%%%%%%%%%%%%%%%%%%

Our AQEC scheme corrects errors induced only by single-photon loss.
Thus, it is important to check how other relaxation channels, photon gain and dephasing, degrade our AQEC scheme.
Figure~\ref{fig:relax}(a) and (b) shows how much the bit- and phase-flip rates increase with thermal photon number $n_\textrm{th}$, which characterizes the photon gain process, and dephasing rate $\gamma_\phi$ in Eq.~\eqref{eq:lindblad}.
Note that photon gain and dephasing contribute differently to population leakage: photon gain increases populations in both the error and HEL spaces, while dephasing induces population leakage to the HEL space only, as shown in Fig.~\ref{fig:relax}(c).

Now, we discuss the upper bounds of $n_\textrm{th}$ and $\gamma_\phi$ for reliable error correction.
According to Fig.~\ref{fig:relax}(a), $n_\textrm{th}$ must be less than 0.01 to keep the increase in the flip rates less than 20\%.
For $\gamma_\phi$, although a four-photon KPO is insensitive to dephasing induced by low-frequency noise as pointed out in Sec.~\ref{sec:sim}, the KPO may be exposed to effective dephasing caused by quantum jumps in a nearby quantum system.
The rate of such a dephasing process is given by \cite{clerk2007, rigetti2012}:
\begin{equation}\label{eq:decay}
\gamma_\phi = 
\frac{\gamma_\textrm{NQS}}{2}
\Re\!\left[ \sqrt{\left(1+\frac{2\textrm{i}\chi}{\gamma_\textrm{NQS}} \right)^2 + \frac{8\textrm{i}\chi}{\gamma_\textrm{NQS}}n_\textrm{th}^\textrm{q}} - 1 \right],
\end{equation}
where 
$\gamma_\textrm{NQS}$ and $n_\textrm{th}^\textrm{q}$ are the damping rate and the thermal photon number of the nearby quantum system, respectively, and
$\chi$ is the dispersive shift between the KPO and the nearby quantum system.
We first consider dephasing induced by the ancilla resonator.
In this case, $\gamma_\textrm{NQS} = \gamma_\textrm{an}$ and $\chi \equiv K[g/(\omega_\textrm{an}-\omega_\textrm{KPO})]^2$.
Even if the thermal photon number of the ancilla is as large as 0.1, $\gamma_\phi/2\pi$ is still less than 1 Hz, which is completely negligible.
Similarly, the partial population of the ancilla resonator during AQEC may be concerning.
The mean photon number for this is approximately 0.02; thus, it is also negligible.

Another possible quantum system that can result in effective dephasing is a transmon or a resonator for readout.
Here, we consider a transmon.
In this case, $\chi \gg \gamma_\textrm{NQS}$;
then Eq.~\eqref{eq:decay} becomes $\gamma_\phi \approx n_\textrm{th}^\textrm{q}\gamma_\textrm{NQS}$ \cite{reagor2016}.
If $T_1$ of the transmon is 20 $\mu$s, $n_\textrm{th}^\textrm{q} = 0.02$ yields $\gamma_\phi/2\pi \approx 160$ Hz, where the bit- and phase-flip rates increase significantly [Fig.~\ref{fig:relax}(b)].
Thus, it is crucial to keep the system as cold as possible so that $n_\textrm{th}^\textrm{q} \ll 0.01$ to maximize the performance of the AQEC scheme \cite{jin2015, wang2019}.

\subsection{Unconditional reset}
\label{sec:reset}

%%%%%%%%%%%%%%%%%%%%%%%%%%%%%%%%
\begin{figure}
\centering
\includegraphics{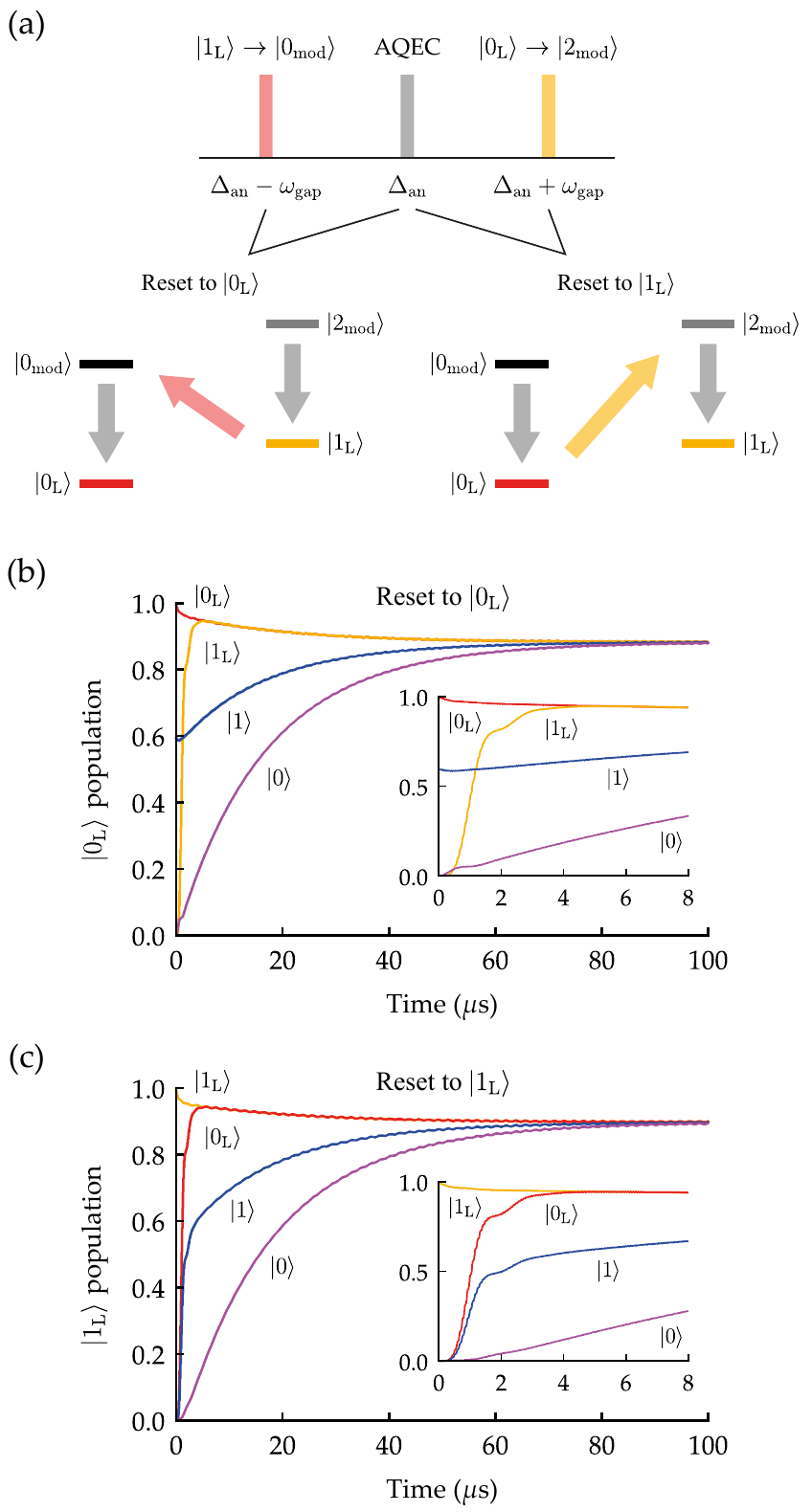}
\caption{Unconditional reset.
(a) Frequencies of the reset/correction tones (upper panel) and population transfer path created by these two tones (lower panel).
The reset tone in red and the correction tone in gray lock the system state to $\ket{0_\textrm{L}}$, whereas the orange and gray lock the system to $\ket{1_\textrm{L}}$ by following the population transfer path.
The frequency of the reset tone is determined by the position of the dip in Fig.~\ref{fig:AQEC}(a).
(b) and (c) Numerical simulation of our reset scheme.
The states near the curves indicate the initial states.
The insets show the short-time behavior.
The parameters for (b) are as follows:
$A_\textrm{cor}/2\pi = 0.50$ MHz,
$\omega_\textrm{reset}/2\pi = \omega_\textrm{cor}/2\pi - 12.2$ MHz,
$A_\textrm{reset}/2\pi = 0.32$ MHz.
The parameters for (c) are as follows:
$A_\textrm{cor}/2\pi = 0.45$ MHz,
$\omega_\textrm{reset}/2\pi = \omega_\textrm{cor}/2\pi + 12.2$ MHz,
$A_\textrm{reset}/2\pi = 0.40$ MHz.
Other parameters are identical to those for Fig.~\ref{fig:AQEC}(b) and (c).
}
\label{fig:reset}
\end{figure}
%%%%%%%%%%%%%%%%%%%%%%%%%%%%%%%%

Our simple AQEC scheme is not the only advantage of a four-photon KPO---now, we introduce an unconditional reset scheme that forces the state of the system to evolve to one of the logical qubit states regardless of the initial state.
In this scheme, an additional microwave tone is required as well as the correction tone.
This additional tone, which we call the reset tone, activates the transitions suppressed by $\omega_\textrm{gap}$ such that all populations within the information space are transferred to either $\ket{0_\textrm{L}}$ or $\ket{1_\textrm{L}}$, depending on the frequency of the reset tone [Fig.~\ref{fig:reset}(a)].
We simulate this scheme by adding the term $\hbar A_\textrm{reset}\cos(\omega_\textrm{reset}t)(\hat{a}^\dagger\hat{b}^\dagger+\hat{a}\hat{b})$ to Eq.~\eqref{eq:full}, where $A_\textrm{reset}$ and $\omega_\textrm{reset}$ are the amplitude and frequency of this term, respectively.

Figure~\ref{fig:reset}(b) shows the population of $\ket{0_\textrm{L}}$ as a function of time when the system is exposed to the correction and the reset tones with $\omega_\textrm{reset} = \Delta_\textrm{an}-\omega_\textrm{gap}$, which locks the system to $\ket{0_\textrm{L}}$.
Note that, the population of $\ket{0_\textrm{L}}$ saturates at about 90\% regardless of the initial state.
Thus, we can reset the logical qubit simply by applying two microwave tones without any state preparation.
If the initial state is in the information space, the KPO state can reach the target state in less than 5 $\mu$s [the inset of Fig.~\ref{fig:reset}(b)];
however, if the initial state is outside of the information space, such as Fock states, the reset might take nearly 100 $\mu$s mainly because of the protection energy gap.
Reset to $\ket{1_\textrm{L}}$ can be carried out by setting $\omega_\textrm{reset} = \Delta_\textrm{an}+\omega_\textrm{gap}$ [Fig.~\ref{fig:reset}(c)].

One may find some similarity between this reset scheme and dynamic nuclear polarization \cite{dnp, slichter}
because the population transfer path in Fig.~\ref{fig:reset}(a) is identical to that of dynamic nuclear polarization in an interacting nucleus--electron pair of spins $1/2$.
Transitions from the code space to the error space correspond to the electron spin excitation and
transitions within the code space correspond to the nuclear spin excitation.
We stress that, however, our scheme is more general than dynamic nuclear polarization because our scheme can reset even a state outside of the information space.

\subsection{Gate operation and circuit implementation}
\label{sec:gate}

%%%%%%%%%%%%%%%%%%%%%%%%%%%%%%%%
\begin{figure}
\centering
\includegraphics{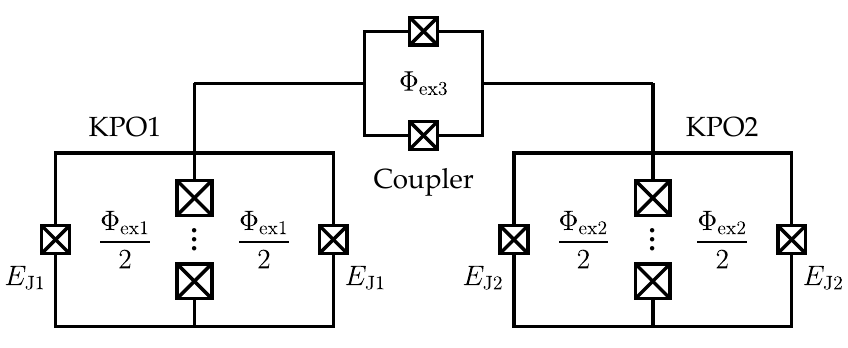}
\caption{Circuit implementation.
Here, two KPOs are coupled via a flux-tunable coupler, which is a direct current superconducting quantum interference device (DC SQUID).
Two large junctions with vertical dots represent a junction arrays.
The junction capacitances and ancilla resonators are not shown for simplicity.
}
\label{fig:circuit}
\end{figure}
%%%%%%%%%%%%%%%%%%%%%%%%%%%%%%%%

Thus far, we have focused on error correction and state preparation.
In this section, we briefly discuss gate operations and circuit implementation.
Since our code relies on the fourfold rotational symmetry, the X gate can be implemented by a two-photon drive with the frequency $\omega_\textrm{p}/2 \pm \omega_\textrm{gap}$ as shown in Figs.~\ref{fig:protect}(b) and \ref{fig:AQEC}(b).
QuTiP simulations for X gate operations are presented in Supplementary Figure 3.
The Z gate can be implemented by waiting for the time $\pi/\omega_\textrm{gap}$ or by shifting the phase of the subsequent drive (virtual Z gate \cite{nmr, mckay2017}).

One possible circuit implementation is shown in Fig.~\ref{fig:circuit}.
Note that we employ two symmetric loops for the KPO.
(A similar circuit was used differently in Ref.~\cite{lescanne2020}.)
Because of this symmetry, no current flows through the junction array.
Thus, we can separate the system into a weakly nonlinear inductor (junction array) and a symmetric DC SQUID.
One consequence of this is that, for KPO1, almost linear modulation of the junction energy can be obtained at $\Phi_\textrm{ex1} = 0.5\Phi_0$, which we call the optimal bias.
The reason is that the effective junction energy of a symmetric DC SQUID is given by $2E_\textrm{J1}\cos(\pi\Phi_\textrm{ex1}/\Phi_0)$ \cite{qubitComp}, which results in $-2E_\textrm{J1}\sin(\pi\Phi_\textrm{ac1}/\Phi_0)$ at the optimal bias, where $\Phi_\textrm{ac1}$ is an oscillating flux passing through KPO1.
Thus, by setting the frequency of $\Phi_\textrm{ac1}$ close to $4\omega_\textrm{KPO}$ and $2\omega_\textrm{KPO}$, we obtain the four-photon pump and the two-photon drive from the five- and three-wave mixing, respectively, without having unwanted processes from even terms of $\Phi_\textrm{ac1}$ (see Supplementary Note 1).
Another consequence is that the Kerr coefficient is mainly determined by the junction array at the optimal bias.
Such a functional separation allows us to design the circuit conveniently.

One potential problem regarding actual experiments is that the resulting amplitude of the four-photon pump [$P$ in Eq.~\eqref{eq:KPO}] might be too small.
We find that $P$ is proportional to $KN^3$ at the optimal bias, where $N$ is the number of Josephson junctions in the junction array (see Supplementary Note 1).
Hence, it is advantageous to select $N \gg 1$.
However, $N$ cannot be arbitrarily large:
the capacitive energy of KPO, which is given by $KN^2$, is limited by an intrinsic capacitive energy of a Josephson junction, which is approximately a few GHz. Thus, if the target $K$ is approximately a few tens of MHz, then $N$ cannot exceed 10.
One may apply a more advanced technique that was originally developed for a dissipative parametric oscillator to generate higher-order nonlinearity from lower-order parametric processes \cite{mundhada2017, mundhada2019}.

To complete a universal gate set, we need a two-qubit gate.
Here, we consider the iSWAP $\hat{U}_\textrm{iSWAP}$ and bSWAP $\hat{U}_\textrm{bSWAP}$ gates, which are defined by \cite{niskanen2007, poletto2012}
\begin{equation}
\hat{U}_\textrm{iSWAP} =
\begin{pmatrix}
1 & 0 & 0 & 0 \\
0 & 0 & -\textrm{i} & 0 \\
0 & -\textrm{i} & 0 & 0 \\
0 & 0 & 0 & 1 \\
\end{pmatrix}, \quad
\hat{U}_\textrm{bSWAP} =
\begin{pmatrix}
0 & 0 & 0 & -\textrm{i} \\
0 & 1 & 0 & 0 \\
0 & 0 & 1 & 0 \\
-\textrm{i} & 0 & 0 & 0 \\
\end{pmatrix}.
\end{equation}
The iSWAP gate requires two-photon exchange terms, i.e., $\hat{a}_1^\dagger\hat{a}_1^\dagger\hat{a}_2\hat{a}_2 + \hat{a}_1\hat{a}_1\hat{a}_2^\dagger\hat{a}_2^\dagger$, where $\hat{a}_i$ and $\hat{a}_i^\dagger$ are the ladder operators for KPO $i$ ($i=1,2$);
the bSWAP gates requires $\hat{a}_1^\dagger\hat{a}_1^\dagger\hat{a}_2^\dagger\hat{a}_2^\dagger + \hat{a}_1\hat{a}_1\hat{a}_2\hat{a}_2$.
These terms can be induced without disturbing the optimal bias of each KPO by applying a parametric drive to a tunable coupler \cite{niskanen2007, niskanen2006}.
In Fig.~\ref{fig:circuit}, a DC SQUID is employed as a tunable coupler.
In such a configuration, a parametric drive with the frequency $2\abs{\omega_\textrm{KPO1}-\omega_\textrm{KPO2}}$, where $\omega_{\textrm{KPO}i}$ is the transition frequency of KPO $i$, gives the two-photon exchange terms at $\Phi_\textrm{ex3} = 0.5\Phi_0$, thus resulting in the iSWAP gate (see Supplementary Note 2).
Similarly, a parametric drive with the frequency $2(\omega_\textrm{KPO1}+\omega_\textrm{KPO2})$ implements the bSWAP gate at the same flux bias.

\section{Discussion}

Here, we list some comments on future research directions.
First, the present work is based on numerical analysis.
Further general and analytic treatment on the physics underlying this scheme is desirable;
in particular, the increase in the phase-flip time due to the detuning of $P$ must be clarified.

Second, a convenient single-shot readout scheme must be developed, such as cat-quadrature readout for cat states in a two-photon KPO \cite{puri2019, grimm2020, pfaff2017}.

Third, the unconditional reset scheme must be improved to enhance the final population of the target logical state.
Since about half of the lost population is in the error space and the other half is in the HEL space (see Supplementary Table 1), we must develop a scheme that transfers the population of the HEL space to the information space without too much cost.

Lastly, a more efficient optimization procedure is required.
In this work, the essential parameters, such as the frequency and amplitude of microwave tones as well as the single-photon loss rate of the ancilla, are determined by sweeping the parameter space as shown in Figs.~\ref{fig:AQEC}(a) and \ref{fig:opt}(c).
One interesting research direction is to combine our schemes and the automation procedure developed in Ref.~\cite{wang2021}.

In summary, we have proposed an AQEC scheme that requires only one continuous microwave tone to correct error autonomously.
This scheme is based on
(i) the protection of the information space by applying a four-photon pump to a KPO,
(ii) the energy degeneracy between $\ket{0_\textrm{mod}}$ and $\ket{1_\textrm{mod}}$ as well as between $\ket{2_\textrm{mod}}$ and $\ket{3_\textrm{mod}}$,
(iii) one-way transition using a lossy ancilla resonator, and
(iv) suppressing unwanted transition by creating an energy gap.
By solving the master equation, we show that the relaxation times of the logical qubit states surpass the break-even point with our AQEC scheme.
In addition to AQEC, we introduce an unconditional reset scheme that lets the system evolve into one of the logical qubit states by simply applying two continuous microwave tones.

Complications in bosonic codes originate from state-by-state control in the Fock basis.
This is a consequence of using the dispersive coupling between a bosonic system and a nonlinear ancilla for control \cite{heeres2015, krastanov2015}.
A four-photon KPO can be a radically different approach because its finite anharmonicity allows us to control the system without an ancilla, and the logical qubit states are quasienergy eigenstates such that AQEC and gate operation do not need to rely on the Fock basis.
This suggests that we can apply the intuition acquired from conventional two-level-system qubits to a four-photon KPO;
the similarity between our reset scheme and dynamic nuclear polarization can be an example of this.
Thus, we believe our AQEC and reset schemes reduce hardware overhead significantly, making a KPO an essential unit for future bosonic quantum computing systems.

\section*{Data availability}
Datasets generated from the simulation are available from the corresponding
authors upon reasonable request.

\begin{acknowledgments}
The authors would like to thank Akiyoshi Tomonaga for helpful discussion and the reviewers for constructive and thoughtful comments.
This work was supported by the Japan Science and Technology Agency (CREST, JPMJCR1676; Moonshot R\&D, JPMJMS2067) and the New Energy and Industrial Technology Development Organization (NEDO, JPNP16007).
\end{acknowledgments}

\section*{Author contributions}
S.K. and J.-S.T conceived the project.
S.K. constructed the schemes, performed the numerical simulations, and wrote the manuscript.
S.K. and S.W. derived the equations in Supplementary Notes.
J.-S.T supervised the project.
All authors edited the paper.

\section*{Competing interests}
The authors declare no competing interests.

\newpage
$ $
\includepdf[pages=1]{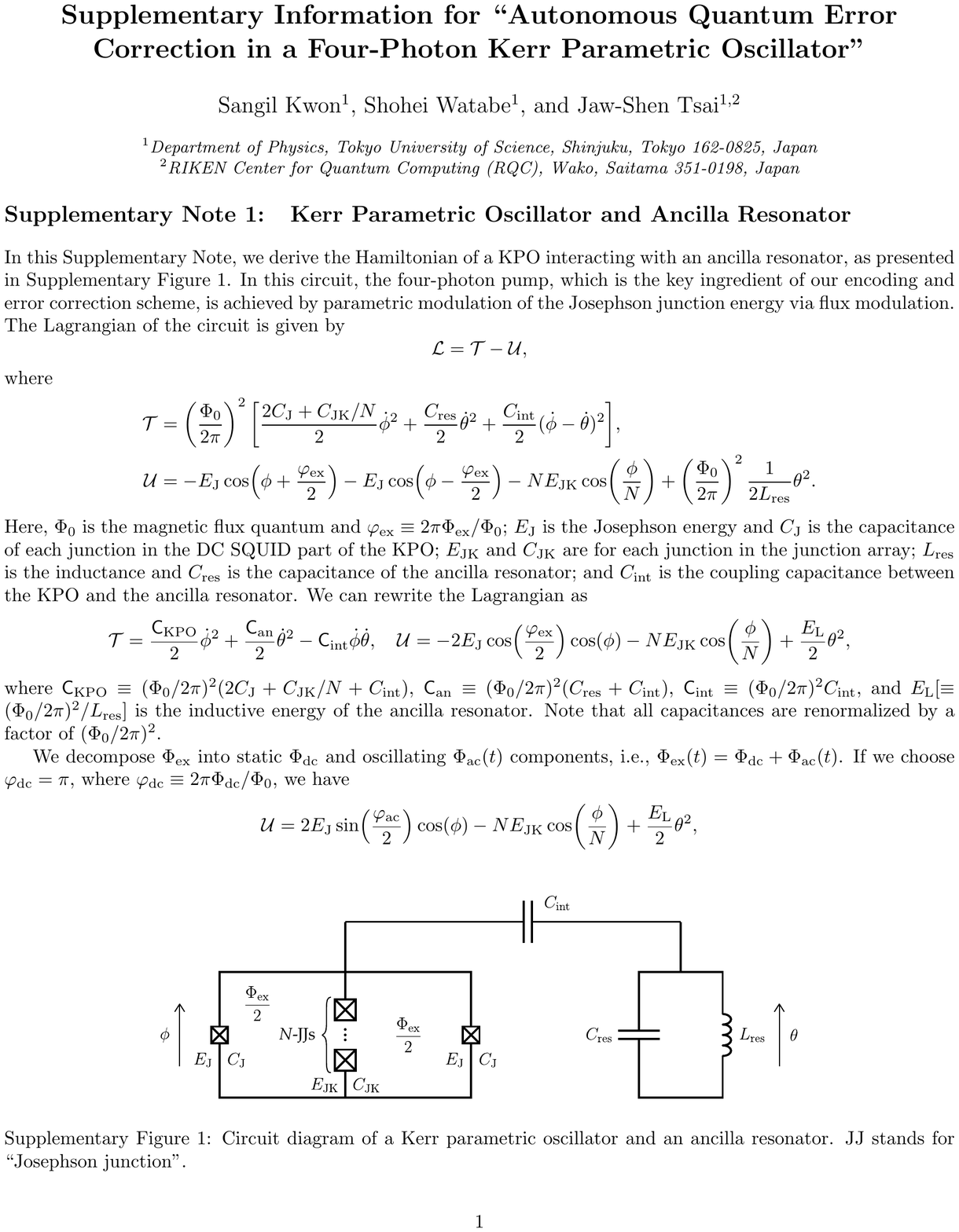}
\newpage
$ $
\includepdf[pages=2]{supp.pdf}
\newpage
$ $
\includepdf[pages=3]{supp.pdf}
\newpage
$ $
\includepdf[pages=4]{supp.pdf}
\newpage
$ $
\includepdf[pages=5]{supp.pdf}
\newpage
$ $


\section*{Supplementary material}

# Supplementary Information for "Autonomous Quantum Error Correction in a Four-Photon Kerr Parametric Oscillator"

Sangil Kwon[1], Shohei Watabe[1], and Jaw-Shen Tsai[1,2]

[1]*Department of Physics, Tokyo University of Science, Shinjuku, Tokyo 162-0825, Japan*
[2]*RIKEN Center for Quantum Computing (RQC), Wako, Saitama 351-0198, Japan*

**Supplementary Note 1: Kerr Parametric Oscillator and Ancilla Resonator**

In this Supplementary Note, we derive the Hamiltonian of a KPO interacting with an ancilla resonator, as presented in Supplementary Figure 1. In this circuit, the four-photon pump, which is the key ingredient of our encoding and error correction scheme, is achieved by parametric modulation of the Josephson junction energy via flux modulation. The Lagrangian of the circuit is given by

$$\mathcal{L} = \mathcal{T} - \mathcal{U},$$

where

$$\mathcal{T} = \left(\frac{\Phi_0}{2\pi}\right)^2 \left[\frac{2C_\mathrm{J} + C_\mathrm{JK}/N}{2}\dot{\phi}^2 + \frac{C_\mathrm{res}}{2}\dot{\theta}^2 + \frac{C_\mathrm{int}}{2}(\dot{\phi} - \dot{\theta})^2\right],$$

$$\mathcal{U} = -E_\mathrm{J}\cos\left(\phi + \frac{\varphi_\mathrm{ex}}{2}\right) - E_\mathrm{J}\cos\left(\phi - \frac{\varphi_\mathrm{ex}}{2}\right) - NE_\mathrm{JK}\cos\left(\frac{\phi}{N}\right) + \left(\frac{\Phi_0}{2\pi}\right)^2 \frac{1}{2L_\mathrm{res}}\theta^2.$$

Here, $\Phi_0$ is the magnetic flux quantum and $\varphi_\mathrm{ex} \equiv 2\pi\Phi_\mathrm{ex}/\Phi_0$; $E_\mathrm{J}$ is the Josephson energy and $C_\mathrm{J}$ is the capacitance of each junction in the DC SQUID part of the KPO; $E_\mathrm{JK}$ and $C_\mathrm{JK}$ are for each junction in the junction array; $L_\mathrm{res}$ is the inductance and $C_\mathrm{res}$ is the capacitance of the ancilla resonator; and $C_\mathrm{int}$ is the coupling capacitance between the KPO and the ancilla resonator. We can rewrite the Lagrangian as

$$\mathcal{T} = \frac{\mathsf{C}_\mathrm{KPO}}{2}\dot{\phi}^2 + \frac{\mathsf{C}_\mathrm{an}}{2}\dot{\theta}^2 - \mathsf{C}_\mathrm{int}\dot{\phi}\dot{\theta}, \quad \mathcal{U} = -2E_\mathrm{J}\cos\left(\frac{\varphi_\mathrm{ex}}{2}\right)\cos(\phi) - NE_\mathrm{JK}\cos\left(\frac{\phi}{N}\right) + \frac{E_\mathrm{L}}{2}\theta^2,$$

where $\mathsf{C}_\mathrm{KPO} \equiv (\Phi_0/2\pi)^2(2C_\mathrm{J} + C_\mathrm{JK}/N + C_\mathrm{int})$, $\mathsf{C}_\mathrm{an} \equiv (\Phi_0/2\pi)^2(C_\mathrm{res} + C_\mathrm{int})$, $\mathsf{C}_\mathrm{int} \equiv (\Phi_0/2\pi)^2 C_\mathrm{int}$, and $E_\mathrm{L}[\equiv (\Phi_0/2\pi)^2/L_\mathrm{res}]$ is the inductive energy of the ancilla resonator. Note that all capacitances are renormalized by a factor of $(\Phi_0/2\pi)^2$.

We decompose $\Phi_\mathrm{ex}$ into static $\Phi_\mathrm{dc}$ and oscillating $\Phi_\mathrm{ac}(t)$ components, i.e., $\Phi_\mathrm{ex}(t) = \Phi_\mathrm{dc} + \Phi_\mathrm{ac}(t)$. If we choose $\varphi_\mathrm{dc} = \pi$, where $\varphi_\mathrm{dc} \equiv 2\pi\Phi_\mathrm{dc}/\Phi_0$, we have

$$\mathcal{U} = 2E_\mathrm{J}\sin\left(\frac{\varphi_\mathrm{ac}}{2}\right)\cos(\phi) - NE_\mathrm{JK}\cos\left(\frac{\phi}{N}\right) + \frac{E_\mathrm{L}}{2}\theta^2,$$

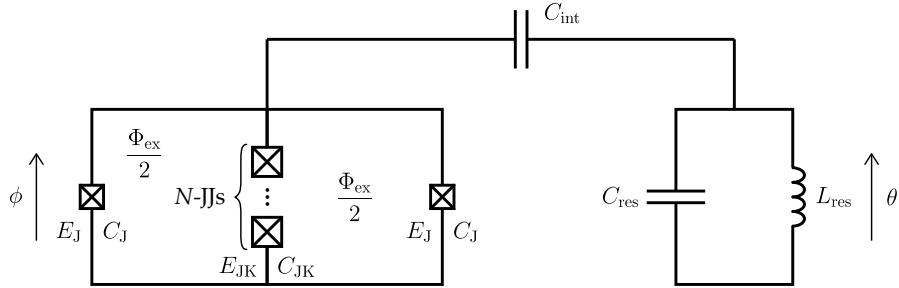

Supplementary Figure 1: Circuit diagram of a Kerr parametric oscillator and an ancilla resonator. JJ stands for "Josephson junction".

where $\varphi_{\text{ac}} \equiv 2\pi\Phi_{\text{ac}}/\Phi_0$. We call the bias $\varphi_{\text{dc}} = \pi$ ($\Phi_{\text{dc}} = 0.5\Phi_0$) as the optimal bias.

The conjugate number operators are defined by

$$\hbar N_\phi = \frac{\partial \mathcal{L}}{\partial \dot\phi} = \mathsf{C}_{\text{KPO}}\dot\phi - \mathsf{C}_{\text{int}}\dot\theta, \quad \hbar N_\theta = \frac{\partial \mathcal{L}}{\partial \dot\theta} = -\mathsf{C}_{\text{int}}\dot\phi + \mathsf{C}_{\text{an}}\dot\theta,$$

where the reduced Planck constants are inserted to make $N_\phi$ and $N_\theta$ dimensionless. In matrix form,

$$\hbar \begin{pmatrix} N_\phi \\ N_\theta \end{pmatrix} = \begin{pmatrix} \mathsf{C}_{\text{KPO}} & -\mathsf{C}_{\text{int}} \\ -\mathsf{C}_{\text{int}} & \mathsf{C}_{\text{an}} \end{pmatrix} \begin{pmatrix} \dot\phi \\ \dot\theta \end{pmatrix}.$$

Using the inverse capacitance matrix, we obtain

$$\begin{pmatrix} \dot\phi \\ \dot\theta \end{pmatrix} = \frac{\hbar}{\mathsf{C}_{\text{KPO}}\mathsf{C}_{\text{an}} - \mathsf{C}_{\text{int}}^2} \begin{pmatrix} \mathsf{C}_{\text{an}} & \mathsf{C}_{\text{int}} \\ \mathsf{C}_{\text{int}} & \mathsf{C}_{\text{KPO}} \end{pmatrix} \begin{pmatrix} N_\phi \\ N_\theta \end{pmatrix}.$$

The resulting Hamiltonian is (from now on, we add a hat to an operator for clarity)

$$\hat{\mathcal{H}} = \hbar^2 \frac{\mathsf{C}_{\text{an}}\hat N_\phi^2 + \mathsf{C}_{\text{KPO}}\hat N_\theta^2 + 2\mathsf{C}_{\text{int}}\hat N_\phi \hat N_\theta}{2(\mathsf{C}_{\text{KPO}}\mathsf{C}_{\text{an}} - \mathsf{C}_{\text{int}}^2)} + 2E_{\text{J}}\sin\left(\frac{\varphi_{\text{ac}}}{2}\right)\cos(\hat\phi) - NE_{\text{JK}}\cos\left(\frac{\hat\phi}{N}\right) + \frac{E_{\text{L}}}{2}\hat\theta^2$$

$$= 4E_{\text{C}}^\phi \hat N_\phi^2 + 4E_{\text{C}}^\theta \hat N_\theta^2 + 8E_{\text{C}}^{\text{int}} \hat N_\phi \hat N_\theta + 2E_{\text{J}}\sin\left(\frac{\varphi_{\text{ac}}}{2}\right)\cos(\hat\phi) - NE_{\text{JK}}\cos\left(\frac{\hat\phi}{N}\right) + \frac{E_{\text{L}}}{2}\hat\theta^2,$$

where

$$E_{\text{C}}^\phi \equiv \frac{\hbar^2 \mathsf{C}_{\text{an}}}{8(\mathsf{C}_{\text{KPO}}\mathsf{C}_{\text{an}} - \mathsf{C}_{\text{int}}^2)}, \quad E_{\text{C}}^\theta \equiv \frac{\hbar^2 \mathsf{C}_{\text{KPO}}}{8(\mathsf{C}_{\text{KPO}}\mathsf{C}_{\text{an}} - \mathsf{C}_{\text{int}}^2)}, \quad E_{\text{C}}^{\text{int}} \equiv \frac{\hbar^2 \mathsf{C}_{\text{int}}}{8(\mathsf{C}_{\text{KPO}}\mathsf{C}_{\text{an}} - \mathsf{C}_{\text{int}}^2)}.$$

We move to the occupation-number representation by defining

$$\hat N_\phi = \mathrm{i} N_0^\phi (\hat a^\dagger - \hat a), \quad \hat\phi = \phi_0 (\hat a^\dagger + \hat a), \quad \hat N_\theta = \mathrm{i} N_0^\theta (\hat b^\dagger - \hat b), \quad \hat\theta = \theta_0 (\hat b^\dagger + \hat b),$$

where $N_0^\phi = \sqrt[4]{E_{\text{JK}}/32NE_{\text{C}}^\phi}$, $\phi_0 = \sqrt[4]{2NE_{\text{C}}^\phi/E_{\text{JK}}}$, $N_0^\theta = \sqrt[4]{E_{\text{L}}/32E_{\text{C}}^\theta}$, and $\theta_0 = \sqrt[4]{2E_{\text{C}}^\theta/E_{\text{L}}}$ are the zero-point fluctuations. Then we have

$$\hat{\mathcal{H}} = \hbar\omega_{\text{KPO}} \hat a^\dagger \hat a + \hbar\omega_{\text{an}} \hat b^\dagger \hat b + \hbar g(\hat a^\dagger \hat b + \hat a \hat b^\dagger - \hat a \hat b - \hat a^\dagger \hat b^\dagger)$$
$$- NE_{\text{JK}}\cos\left(\frac{\hat\phi}{N}\right) - \frac{E_{\text{JK}}}{2N}\hat\phi^2 + 2E_{\text{J}}\sin\left(\frac{\varphi_{\text{ac}}}{2}\right)\cos(\hat\phi), \tag{1}$$

where

$$\hbar\omega_{\text{KPO}} \equiv \sqrt{8E_{\text{C}}^\phi E_{\text{JK}}/N}, \quad \hbar\omega_{\text{an}} \equiv \sqrt{8E_{\text{C}}^\theta E_{\text{L}}}, \quad \hbar g \equiv E_{\text{C}}^{\text{int}} \sqrt[4]{\frac{4E_{\text{JK}}E_{\text{L}}}{NE_{\text{C}}^\phi E_{\text{C}}^\theta}}.$$

The Taylor series for Eq. (1) is

$$\hat{\mathcal{H}}_{\text{JJ}} = -NE_{\text{JK}}\left[1 + \frac{1}{24}\left(\frac{\hat\phi}{N}\right)^4 - \cdots\right] + 2E_{\text{J}}\sin\left(\frac{\varphi_{\text{ac}}}{2}\right)\left(1 - \frac{\hat\phi^2}{2} + \frac{\hat\phi^4}{24} - \cdots\right).$$

After normal ordering, we obtain

$$-\frac{E_{\text{JK}}}{24N^3}\hat\phi^4 \quad \Rightarrow \quad -\hbar K \hat a^\dagger \hat a - \frac{\hbar K}{2} \hat a^\dagger \hat a^\dagger \hat a \hat a$$
$$-\frac{\hbar K}{12}\left(6\hat a^\dagger \hat a^\dagger + 6\hat a \hat a + 4\hat a^\dagger \hat a \hat a + 4\hat a^\dagger \hat a^\dagger \hat a^\dagger \hat a + \hat a \hat a \hat a \hat a + \hat a^\dagger \hat a^\dagger \hat a^\dagger \hat a^\dagger\right),$$

$$-E_{\text{J}}\sin\left(\frac{\varphi_{\text{ac}}}{2}\right)\hat\phi^2 \quad \Rightarrow \quad -\frac{\hbar\omega_{\text{KPO}}N}{2\gamma}\sin\left(\frac{\varphi_{\text{ac}}}{2}\right)\left(2\hat a^\dagger \hat a + \hat a^\dagger \hat a^\dagger + \hat a \hat a\right),$$

$$\frac{E_{\text{J}}}{12}\sin\left(\frac{\varphi_{\text{ac}}}{2}\right)\hat\phi^4 \quad \Rightarrow \quad \frac{2\hbar KN^3}{\gamma}\sin\left(\frac{\varphi_{\text{ac}}}{2}\right)\hat a^\dagger \hat a + \frac{\hbar KN^3}{\gamma}\sin\left(\frac{\varphi_{\text{ac}}}{2}\right)\hat a^\dagger \hat a^\dagger \hat a \hat a$$
$$+ \frac{\hbar KN^3}{6\gamma}\sin\left(\frac{\varphi_{\text{ac}}}{2}\right)\left(6\hat a^\dagger \hat a^\dagger + 6\hat a \hat a + 4\hat a^\dagger \hat a \hat a + 4\hat a^\dagger \hat a^\dagger \hat a^\dagger \hat a + \hat a \hat a \hat a \hat a + \hat a^\dagger \hat a^\dagger \hat a^\dagger \hat a^\dagger\right),$$

where $\hbar K \equiv E_C^\phi/N^2$ and $\gamma \equiv E_{JK}/E_J$. We stress that the transition frequency of the KPO is $\omega_{KPO} - K$, whereas the same quantity is denoted as $\omega_{KPO}$ in the main text.

Note that we apply two parametric tones: one is the four-photon pump with the frequency $\omega_p$ and the other is the two-photon drive with the frequency $\omega_d$. Since the amplitudes of these two tones are usually small, we can approximate $\sin[\varphi_{ac}(t)]$ as $\varphi_p(t) = \varphi_{p0}\cos(\omega_p t)$ for the pump and $\varphi_d(t) = \varphi_{d0}\cos(\omega_d t)$ for the drive.

Now we move to the rotating frame whose Hamiltonian is defined by

$$\hat{\mathcal{H}}_0 = \hbar\frac{\omega_p}{4}\left(\hat{a}^\dagger\hat{a} + \hat{b}^\dagger\hat{b}\right)$$

and take the rotating wave approximation. Then the surviving terms are

$$\hat{\mathcal{H}}_{rot} = e^{i\hat{\mathcal{H}}_0 t/\hbar}(\hat{\mathcal{H}} - \hat{\mathcal{H}}_0)e^{-i\hat{\mathcal{H}}_0 t/\hbar}$$
$$\approx \underbrace{\hbar\Delta_{KPO}\hat{a}^\dagger\hat{a} + \hbar\Delta_{an}\hat{b}^\dagger\hat{b} + \hbar g(\hat{a}^\dagger\hat{b} + \hat{a}\hat{b}^\dagger)}_{\text{linear part, }\hat{\mathcal{H}}_{lin}} \underbrace{- \hbar\frac{K}{2}\hat{a}^\dagger\hat{a}^\dagger\hat{a}\hat{a} + \hbar\frac{P}{2}\left(\hat{a}^\dagger\hat{a}^\dagger\hat{a}^\dagger\hat{a}^\dagger + \hat{a}\hat{a}\hat{a}\hat{a}\right) + \hbar\frac{G}{2}\left(\hat{a}^\dagger\hat{a}^\dagger + \hat{a}\hat{a}\right)}_{\text{nonlinear part, }\hat{\mathcal{H}}_{nl}} \quad (2)$$

where $\Delta_{KPO}(\equiv \omega_{KPO} - K - \omega_p/4)$ is the KPO-pump frequency detuning, $\Delta_{an}(\equiv \omega_{an} - \omega_p/4)$ is the ancilla-pump frequency detuning, $P(\equiv \varphi_{p0} KN^3/12\gamma)$ is the amplitude of the four-photon pump, and $G(\equiv -\varphi_{d0}\omega_{KPO}N/4\gamma)$ is the amplitude of the two-photon drive. Equation (2) is Eq. (4) in the main text with the two-photon drive. Since $P$ is proportional to $KN^3$, choosing $N \gg 1$ ensures large $P$ as pointed out in the main text.

We first diagonalize the linear terms by the Schrieffer–Wolff transformation [1]. The unitary operator for this $\hat{U}_{SW}$ is given by

$$\hat{U}_{SW} = \exp\left[\Lambda(\hat{a}\hat{b}^\dagger - \hat{a}^\dagger\hat{b})\right].$$

This gives

$$\hat{U}_{SW}\hat{a}\hat{U}_{SW}^\dagger = \cos(\Lambda)\hat{a} + \sin(\Lambda)\hat{b}, \quad \hat{U}_{SW}\hat{b}\hat{U}_{SW}^\dagger = -\sin(\Lambda)\hat{a} + \cos(\Lambda)\hat{b}. \quad (3)$$

Using Eq. (3), we obtain

$$\hat{\mathcal{H}}'_{lin} = \hat{U}_{SW}\hat{\mathcal{H}}_{lin}\hat{U}_{SW}^\dagger$$
$$= \hbar\tilde{\Delta}_{KPO}\hat{a}^\dagger\hat{a} + \hbar\tilde{\Delta}_{an}\hat{b}^\dagger\hat{b} + \hbar\left[g\cos(2\Lambda) - \frac{\Delta_{anK}}{2}\sin(2\Lambda)\right](\hat{a}\hat{b}^\dagger + \hat{a}^\dagger\hat{b}), \quad (4)$$

where $\Delta_{anK} \equiv \Delta_{an} - \Delta_{KPO}$ and

$$\tilde{\Delta}_{KPO} \equiv \Delta_{KPO}\cos^2(\Lambda) + \Delta_{an}\sin^2(\Lambda) - g\sin(2\Lambda), \quad \tilde{\Delta}_{an} \equiv \Delta_{an}\cos^2(\Lambda) + \Delta_{KPO}\sin^2(\Lambda) + g\sin(2\Lambda).$$

The last term in Eq. (4) can be canceled out by taking $\Lambda = (1/2)\arctan(2\lambda)$ with $\lambda = g/\Delta_{anK}$. Then, Eq. (4) becomes

$$\hat{\mathcal{H}}'_{lin} = \hbar\tilde{\Delta}_{KPO}\hat{a}^\dagger\hat{a} + \hbar\tilde{\Delta}_{an}\hat{b}^\dagger\hat{b},$$

where

$$\tilde{\Delta}_{KPO} = \frac{1}{2}\left(\Delta_{KPO} + \Delta_{an} - \sqrt{\Delta_{anK}^2 + 4g^2}\right), \quad \tilde{\Delta}_{an} = \frac{1}{2}\left(\Delta_{KPO} + \Delta_{an} + \sqrt{\Delta_{anK}^2 + 4g^2}\right).$$

By taking the same transformation on $\hat{\mathcal{H}}_{nl}$, we obtain

$$\hat{\mathcal{H}}'_{nl} = \hat{U}_{SW}\hat{\mathcal{H}}_{nl}\hat{U}_{SW}^\dagger$$
$$= -\hbar\frac{K}{2}\Big[\cos^4(\Lambda)\hat{a}^\dagger\hat{a}^\dagger\hat{a}\hat{a} + \sin^4(\Lambda)\hat{b}^\dagger\hat{b}^\dagger\hat{b}\hat{b} + \cos^2(\Lambda)\sin^2(\Lambda)\left(4\hat{a}^\dagger\hat{a}\hat{b}^\dagger\hat{b} + \hat{a}^\dagger\hat{a}^\dagger\hat{b}\hat{b} + \hat{a}\hat{a}\hat{b}^\dagger\hat{b}^\dagger\right)$$
$$+ 2\cos^3(\Lambda)\sin(\Lambda)\left(\hat{a}^\dagger\hat{a}\hat{a}\hat{b}^\dagger + \hat{a}^\dagger\hat{a}^\dagger\hat{a}\hat{b}\right) + 2\cos(\Lambda)\sin^3(\Lambda)\left(\hat{a}^\dagger\hat{b}^\dagger\hat{b}\hat{b} + \hat{a}\hat{b}^\dagger\hat{b}^\dagger\hat{b}\right)\Big]$$
$$+ \hbar\frac{P}{2}\Big[\cos^4(\Lambda)\left(\hat{a}^\dagger\hat{a}^\dagger\hat{a}^\dagger\hat{a}^\dagger + \hat{a}\hat{a}\hat{a}\hat{a}\right) + \sin^4(\Lambda)\left(\hat{b}^\dagger\hat{b}^\dagger\hat{b}^\dagger\hat{b}^\dagger + \hat{b}\hat{b}\hat{b}\hat{b}\right) + 6\cos^2(\Lambda)\sin^2(\Lambda)\left(\hat{a}^\dagger\hat{a}^\dagger\hat{b}^\dagger\hat{b}^\dagger + \hat{a}\hat{a}\hat{b}\hat{b}\right)$$
$$+ 4\cos^3(\Lambda)\sin(\Lambda)\left(\hat{a}^\dagger\hat{a}^\dagger\hat{a}^\dagger\hat{b}^\dagger + \hat{a}\hat{a}\hat{a}\hat{b}\right) + 4\cos(\Lambda)\sin^3(\Lambda)\left(\hat{a}^\dagger\hat{b}^\dagger\hat{b}^\dagger\hat{b}^\dagger + \hat{a}\hat{b}\hat{b}\hat{b}\right)\Big]$$
$$+ \hbar\frac{G}{2}\Big[\cos^2(\Lambda)\left(\hat{a}^\dagger\hat{a}^\dagger + \hat{a}\hat{a}\right) + \sin^2(\Lambda)\left(\hat{b}^\dagger\hat{b}^\dagger + \hat{b}\hat{b}\right) + 2\cos(\Lambda)\sin(\Lambda)\left(\hat{a}^\dagger\hat{b}^\dagger + \hat{a}\hat{b}\right)\Big].$$

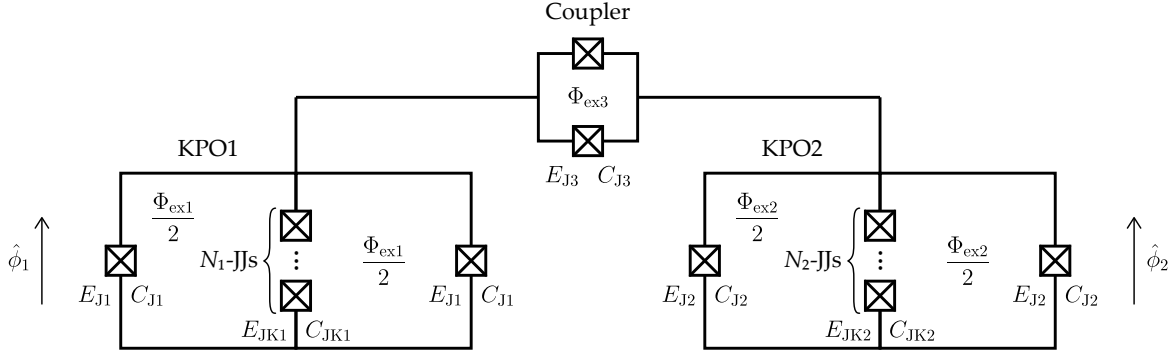

Supplementary Figure 2: Two KPOs coupled by a tunable coupler, which is a symmetric DC SQUID.

In the dispersive limit $\lambda \ll 1$, $\Lambda \approx \lambda$. This gives

$$\hat{\mathcal{H}}'_{\text{lin}} + \hat{\mathcal{H}}'_{\text{nl}} \approx \hbar \left(\Delta_{\text{KPO}} - \Delta_{\text{anK}}\lambda^2\right) \hat{a}^\dagger \hat{a} + \hbar \left(\Delta_{\text{an}} + \Delta_{\text{anK}}\lambda^2\right) \hat{b}^\dagger \hat{b}$$
$$- \hbar \frac{K}{2} \left[ \hat{a}^\dagger \hat{a}^\dagger \hat{a}\hat{a} + 2\lambda \left(\hat{a}^\dagger \hat{a}\hat{a}\hat{b}^\dagger + \hat{a}^\dagger \hat{a}^\dagger \hat{a}\hat{b}\right) + \lambda^2 \left(4\hat{a}^\dagger \hat{a}\hat{b}^\dagger \hat{b} + \hat{a}^\dagger \hat{a}^\dagger \hat{b}\hat{b} + \hat{a}\hat{a}\hat{b}^\dagger \hat{b}^\dagger\right) \right]$$
$$+ \hbar \frac{P}{2} \left[ \left(\hat{a}^\dagger \hat{a}^\dagger \hat{a}^\dagger \hat{a}^\dagger + \hat{a}\hat{a}\hat{a}\hat{a}\right) + 4\lambda \left(\hat{a}^\dagger \hat{a}^\dagger \hat{a}^\dagger \hat{b}^\dagger + \hat{a}\hat{a}\hat{a}\hat{b}\right) + 6\lambda^2 \left(\hat{a}^\dagger \hat{a}^\dagger \hat{b}^\dagger \hat{b}^\dagger + \hat{a}\hat{a}\hat{b}\hat{b}\right) \right]$$
$$+ \hbar \frac{G}{2} \left[ \left(\hat{a}^\dagger \hat{a}^\dagger + \hat{a}\hat{a}\right) + 2\lambda \left(\hat{a}^\dagger \hat{b}^\dagger + \hat{a}\hat{b}\right) + \lambda^2 \left(\hat{b}^\dagger \hat{b}^\dagger + \hat{b}\hat{b}\right) \right].$$

We can see that the amplitude of the AQEC term $A_{\text{cor}}$ is given by $G\lambda$. Since $G/2\pi$ can easily be a few tens of MHz [2, 3], $g/2\pi = 7$ MHz with $\Delta_{\text{anK}}/2\pi = 1$ GHz is sufficient to generate $A_{\text{cor}}$ in the main text. Regarding the dispersive coupling term, its strength is given by $2K\lambda^2/2\pi = 1.9$ kHz, which is more than two orders of magnitude less than $\gamma_{\text{an}}/2\pi = 560$ kHz in the main text.

## Supplementary Note 2: Two Interacting Kerr Parametric Oscillators

In this Supplementary Note, we consider two interacting KPOs via a tunable coupler. For simplicity, we assume that each KPO is perfectly symmetric ($d_J = 0$) and at its optimal bias. Moreover, the coupler is also symmetric and at its optimal bias, i.e., the static component of $\Phi_{\text{ex3}}$ is $0.5\Phi_0$. We use the Hamiltonian of the KPO in Supplementary Note 1 following the notations in Supplementary Figure 2.

$$\hat{\mathcal{H}} = 4E_C^{\phi 1} \hat{N}_{\phi 1}^2 + 4E_C^{\phi 2} \hat{N}_{\phi 2}^2 + 8E_C^{\text{cpl}} \hat{N}_{\phi 1} \hat{N}_{\phi 2} + 2E_{J3} \sin\left(\frac{\varphi_{\text{ac3}}}{2}\right) \cos(\hat{\phi}_1 - \hat{\phi}_2)$$
$$+ 2E_{J1} \sin\left(\frac{\varphi_{\text{ac1}}}{2}\right) \cos(\hat{\phi}_1) - N_1 E_{\text{JK1}} \cos\left(\frac{\hat{\phi}_1}{N_1}\right) + 2E_{J2} \sin\left(\frac{\varphi_{\text{ac2}}}{2}\right) \cos(\hat{\phi}_2) - N_2 E_{\text{JK2}} \cos\left(\frac{\hat{\phi}_2}{N_2}\right).$$

Here, $\varphi_{\text{ac}i}(t) \equiv 2\pi\Phi_{\text{ac}i}(t)/\Phi_0$, where $\Phi_{\text{ac}i}$ is an oscillating flux bias for the KPOs ($i = 1, 2$) and the DC SQUID ($i = 3$). In addition,

$$E_C^{\phi 1} \equiv \frac{\hbar^2 \mathsf{C}_{K2}}{8(\mathsf{C}_{K1}\mathsf{C}_{K2} - \mathsf{C}_{J3}^2)}, \quad E_C^{\phi 2} \equiv \frac{\hbar^2 \mathsf{C}_{K1}}{8(\mathsf{C}_{K1}\mathsf{C}_{K2} - \mathsf{C}_{J3}^2)}, \quad E_C^{\text{cpl}} \equiv \frac{\hbar^2 \mathsf{C}_{J3}}{8(\mathsf{C}_{K1}\mathsf{C}_{K2} - \mathsf{C}_{J3}^2)},$$

where $\mathsf{C}_{K1} \equiv 2\mathsf{C}_{J1} + \mathsf{C}_{JK1}/N_1 + \mathsf{C}_{J3}$ and $\mathsf{C}_{K2} \equiv 2\mathsf{C}_{J2} + \mathsf{C}_{JK2}/N_2 + \mathsf{C}_{J3}$. All capacitances are renormalized by a factor of $(\Phi_0/2\pi)^2$.

After the rotating wave approximation with an assumption that $\varphi_{\text{ac3}}(t) = \varphi_{\text{ac3}}^{(0)} \cos(\omega_{\text{ac3}}t)$, where $\omega_{\text{ac3}} = 2|\omega_{\text{KPO1}} - \omega_{\text{KPO2}}|$, the resulting parametric coupling term $\hat{\mathcal{H}}_{\text{para}}$ is given by

$$\hat{\mathcal{H}}_{\text{para}} = \hbar \frac{\varphi_{\text{ac3}}^{(0)} E_{J3}}{4} \sqrt{\frac{K_1 K_2 N_1^3 N_2^3}{E_{\text{JK1}} E_{\text{JK2}}}} \left(\hat{a}_1^\dagger \hat{a}_1^\dagger \hat{a}_2 \hat{a}_2 + \hat{a}_1 \hat{a}_1 \hat{a}_2^\dagger \hat{a}_2^\dagger\right),$$

where $\hat{a}_i$ and $\hat{a}_i^\dagger$ are the ladder operators for KPO $i$, and $\hbar K_i \equiv E_C^{\phi i}/N_i^2$ ($i = 1, 2$).

4

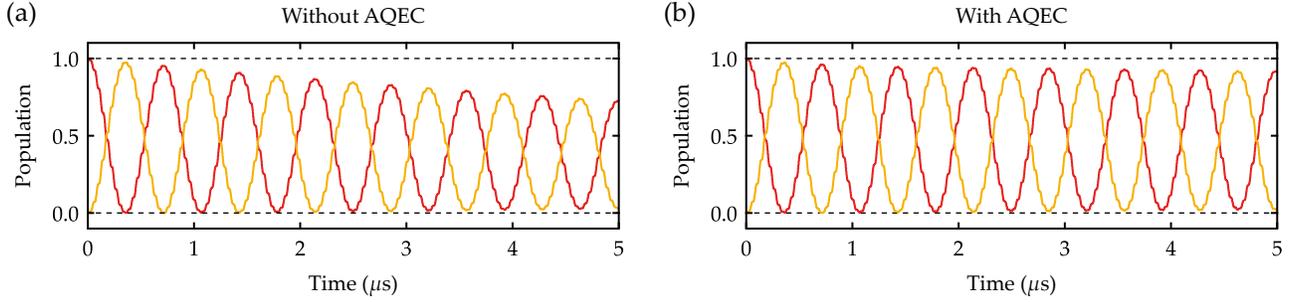

Supplementary Figure 3: X gate with and without AQEC. The red (orange) lines indicate the population of $|0_\text{L}\rangle$ ($|1_\text{L}\rangle$). The amplitude and frequency of the two-photon drive is 0.2 MHz and $\omega_\text{p}/2+\omega_\text{gap}$, respectively. The resulting Rabi frequency $\Omega_\text{Rabi}$ is 1.4 MHz. The frequency of small ripples is $2\omega_\text{gap} \pm \Omega_\text{Rabi}$. Other parameters are the same as those in Fig. 2(b) and (c) in the main text.

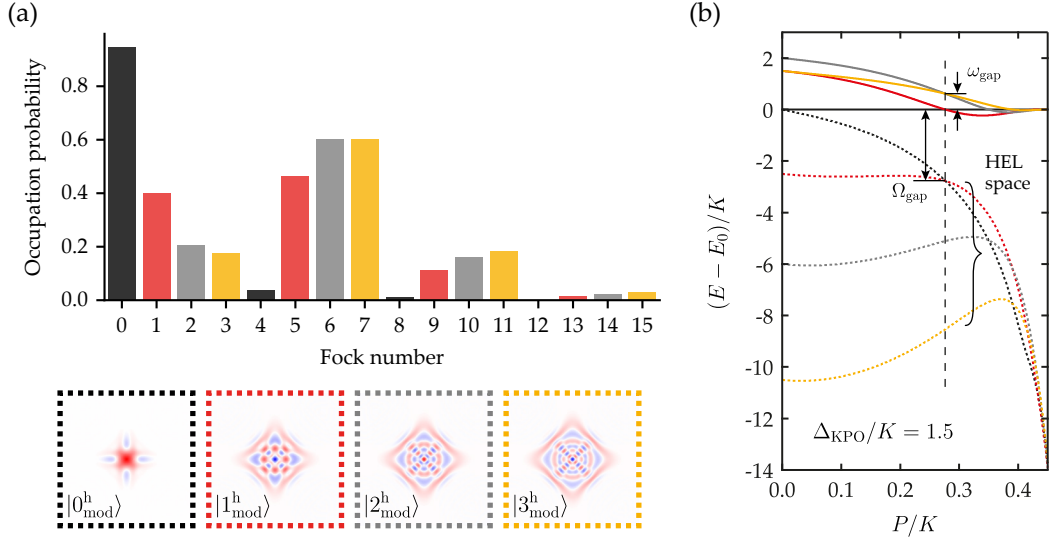

Supplementary Figure 4: (a) Four eigenstates in the HEL space. The upper plot shows the occupation probability of these eigenstates in the Fock basis at $\Delta_{\mathrm{KPO}}/K = 1.5$ and $P/K = 0.2764$ [indicated by the vertical dashed line in (b)]. The four lower plots show the Wigner distribution of these four states. The superscript "h" indicates that the states are in the HEL space. (b) Quasienergy levels of all states considered in this work as a function of $P$. Solid lines are the levels in the information space; dotted lines are the HEL space. The dashed line indicates $P/K = 0.2764$, where the energy levels of $|0_{\mathrm{mod}}\rangle$ and $|1_{\mathrm{mod}}\rangle$ ($|2_{\mathrm{mod}}\rangle$ and $|3_{\mathrm{mod}}\rangle$) degenerate. Here, $\Omega_{\mathrm{gap}}$ indicates the protection energy gap. Colors in this figure indicate the modulus of 4 in the Fock basis.

Supplementary Table 1: Populations of all states considered in this work. The evolution time is 100 $\mu$s, except for the column "Steady state"; the states in parentheses indicate the initial states. The parameters for AQEC and reset are the same as those in Fig. 2 of the main text.

| States | Steady state | AQEC ($|0_{\mathrm{L}}\rangle$) | AQEC ($|1_{\mathrm{L}}\rangle$) | Reset to $|0_{\mathrm{L}}\rangle$ ($|0_{\mathrm{L}}\rangle$) | Reset to $|1_{\mathrm{L}}\rangle$ ($|1_{\mathrm{L}}\rangle$) |
|---|---|---|---|---|---|
| $|0_{\mathrm{mod}}\rangle$ | 0.145 | 0.030 | 0.017 | 0.026 | 0.006 |
| $|1_{\mathrm{mod}}\rangle$ $(=|0_{\mathrm{L}}\rangle)$ | 0.233 | 0.614 | 0.317 | 0.884 | 0.007 |
| $|2_{\mathrm{mod}}\rangle$ | 0.226 | 0.014 | 0.029 | 0.002 | 0.034 |
| $|3_{\mathrm{mod}}\rangle$ $(=|1_{\mathrm{L}}\rangle)$ | 0.181 | 0.264 | 0.572 | 0.005 | 0.897 |
| $|0^{\mathrm{h}}_{\mathrm{mod}}\rangle$ | 0.185 | 0.005 | 0.003 | 0.003 | 0.001 |
| $|1^{\mathrm{h}}_{\mathrm{mod}}\rangle$ | 0.012 | 0.052 | 0.040 | 0.061 | 0.032 |
| $|2^{\mathrm{h}}_{\mathrm{mod}}\rangle$ | 0.009 | 0.010 | 0.014 | 0.008 | 0.019 |
| $|3^{\mathrm{h}}_{\mathrm{mod}}\rangle$ | 0.005 | 0.005 | 0.003 | 0.006 | 0.002 |
| Code space | 0.413 | 0.878 | 0.889 | 0.889 | 0.904 |
| Error space | 0.372 | 0.044 | 0.046 | 0.028 | 0.040 |
| HEL space | 0.212 | 0.073 | 0.061 | 0.077 | 0.053 |
| Total | 0.997 | 0.995 | 0.996 | 0.994 | 0.997 |



\end{document}